\documentclass[twocolumn,
 reprint,
 superscriptaddress,
 amsmath,amssymb,
 aps,
 prl
]{revtex4-1}

\usepackage{graphicx}
\usepackage{dcolumn}
\usepackage{bm}
\usepackage[hidelinks,colorlinks=true,linkcolor=blue,citecolor=blue]{hyperref}
\usepackage[mathlines]{lineno}

\usepackage{braket}
\usepackage{color}
\usepackage{float}
\usepackage{amsmath}
\DeclareMathOperator{\Tr}{Tr}

\begin{document}

\title{Absence of Correlations in Dissipative Interacting Qubits: a No-Go Theorem}

\author{Zeqing Wang}
\affiliation{Department of Physics, Renmin University of China, Beijing, 100872, China}
\affiliation{Shenzhen Institute for Quantum Science and Engineering (SIQSE), Southern University of Science and Technology, Shenzhen, P. R. China.}
\affiliation{International Quantum Academy, Shenzhen 518048, China.}
\affiliation{Guangdong Provincial Key Laboratory of Quantum Science and Engineering, Southern University of Science and Technology Shenzhen, 518055, China.}
\author{Ran Qi}
\affiliation{Department of Physics, Renmin University of China, Beijing, 100872, China}

\author{Yao Lu} 
\affiliation{Shenzhen Institute for Quantum Science and Engineering (SIQSE), Southern University of Science and Technology, Shenzhen, P. R. China.}
\affiliation{International Quantum Academy, Shenzhen 518048, China.}
\affiliation{Guangdong Provincial Key Laboratory of Quantum Science and Engineering, Southern University of Science and Technology Shenzhen, 518055, China.}


\author{Zhigang Wu} 
\email{wuzg@sustech.edu.cn}
\affiliation{Shenzhen Institute for Quantum Science and Engineering (SIQSE), Southern University of Science and Technology, Shenzhen, P. R. China.}
\affiliation{International Quantum Academy, Shenzhen 518048, China.}
\affiliation{Guangdong Provincial Key Laboratory of Quantum Science and Engineering, Southern University of Science and Technology Shenzhen, 518055, China.}
 
\author{Jianwen Jie}
 \email{Jianwen.Jie1990@gmail.com}
\affiliation{Shenzhen Institute for Quantum Science and Engineering (SIQSE), Southern University of Science and Technology, Shenzhen, P. R. China.}
\affiliation{International Quantum Academy, Shenzhen 518048, China.}
\affiliation{Guangdong Provincial Key Laboratory of Quantum Science and Engineering, Southern University of Science and Technology Shenzhen, 518055, China.}

\date{\today}

\begin{abstract}
Exact solutions of model problems are elusive but potent tools for understanding many body interacting systems. We study a system of dissipative qubits with the Heisenberg interaction and obtain, for qubits under a certain condition, an exact steady state solution to the Lindblad master equation describing its dynamics. The physical content of such a solution is a remarkable no-go theorem, which states that for qubits possessing identical ratios of the damping and gain rates, no correlation can be established between them in the steady state. Two consequences of this theorem are discussed in the context of quantum synchronization of qubits. The first is a complete blockade of quantum synchronization of qubits under the aforementioned condition, an effect reminiscent of, but having a much broader scope than, that found in dissipated Kerr-anharmonic oscillators.  The second, and a more important consequence is the possibility of reducing a complex all-to-all qubit network to a much simpler one-to-all network by engineering the dissipation. Such a reduction is desired because it provides an effective tool to optimize the quantum synchronization of a complex qubit network. Finally, we propose two concrete experimental schemes to implement our model and to test our predictions. 
\end{abstract}

\maketitle
{\em Introduction.--- }{\color{black} {\color{black} Understanding the effects of dissipation in an open quantum system caused by its couplings to the environment is one of the most important and urgent tasks in quantum computation and quantum information \cite{breuer2002theory,rivas2011open,RMP2017open,preskill2022}.} {\color{black}Normally, dissipation will {\color{black}{against}} quantum coherence and limit  quantum advantages. Various methods, such as the dynamical decoupling \cite{PRA1998DD,PRL1999DD}, application of high magnetic fields \cite{takahashi2011decoherence} and electromagnetically induced transparency \cite{PRL2000EIT,PRL2001EIT,PRA2002EIT}, have been developed to reduce uncontrolled noises and improve the coherence time for quantum memory~\cite{OQM2009,Freer_2017}. However, suppressing dissipation is often a challenging task in some cases even with the help of these methods. For instance,  reducing the decoherence of the spin ensembles with strong anisotropic interactions remains difficult~\cite{PRL2021DD}.} 

Thus, alternative approaches have been proposed in which dissipation is not suppressed but is carefully designed to facilitate certain tasks of quantum computation~\cite{Harrington2022,PRL1996Engineering}.  This way of treating the dissipation has received increasing attention recently due to rapid progress made in the controllability of quantum devices~\cite{RMP2008control,Nature2010control,Siddiqi2021}.  One notable example of dissipation engineering is the quantum Zeno effect \cite{PRA1990QZE,PRL2017QZE}, which can be used to create a decoherence-free subspace \cite{PRL2000DF,Science2015DF}, an essential ingredient for quantum error correction~\cite{preskill2022,PRA1998QEC,PRA1985QEC,campagne2020quantum,de2022error}. Other examples include preparing maximally entangled states~\cite{lin2013dissipative,shankar2013autonomously,PRL2018QRe,PRL2022OP}, steering quantum states \cite{PRR2020QS,PRL2014QS} and inducing strong multi-body interaction~\cite{PRL20093b,PRL20103b}. Dissipation rate is even proposed as a sensitive probe for monitoring the magnetic noise in nitrogen-vacancy center system~\cite{maze2008nanoscale,zu2021emergent}. All these 
developments are indications that dissipation engineering has tremendous potential of application. 

In this work we investigate dissipation engineering in a model that is of vital importance to both condensed matter physics and to quantum computation, namely a collection of Heisenberg {\color{black}spins} or qubits under both gain and {\color{black}damping}. We employ the widely used Lindblad master equation to describe the dynamics of this system and, for the purpose of exploring the outcomes of the dissipation engineering, focus on its {\color{black} steady states} solutions. Unfortunately, for an interacting system these solutions are difficult to obtain numerically beyond those with a few qubits \cite{RMP2021open}. Exact solutions are also very rare and are limited to certain 1D integrable systems~\cite{Golinelli_2006Exactsol,PRL2011Exact,PRL2013Exactsol,PRL2014Exactsol,PRL2017driven,PRL2020Exactsol}. We thus ask if any general, rigorous statements can be made about the steady states in such an interacting qubit system when the gain and damping of the qubits can be arbitrarily engineered and, if so, whether there are any useful applications for them.

We address these questions by first proving a no-go theorem for the dissipated qubit system. Specifically we argue that it is not possible to create {\color{black}any correlation} between the qubits in the steady state if the qubits share the same ratio of the gain and damping rates. We prove this no-go theorem by showing that under this condition the exact steady state of the interacting system is actually the same as that of the non-interacting system, therefore unable to encode any correlation between the qubits. We further examine the application of this theorem in quantum synchronization of qubits, due to the importance of both dissipation and spin correlation in this phenomenon. We find that in the first place the theorem provides a perfect explanation to an intriguing blockade effect found in the quantum synchronization of qubits. More importantly, it supplies a powerful physical insight in solving the problem of optimizing quantum synchronization of complex qubit networks by dissipation engineering. Finally, we propose two experimental systems in which our no-go theorem and its implications can be tested.

\begin{figure}
\centering
\includegraphics[width=8.3cm]{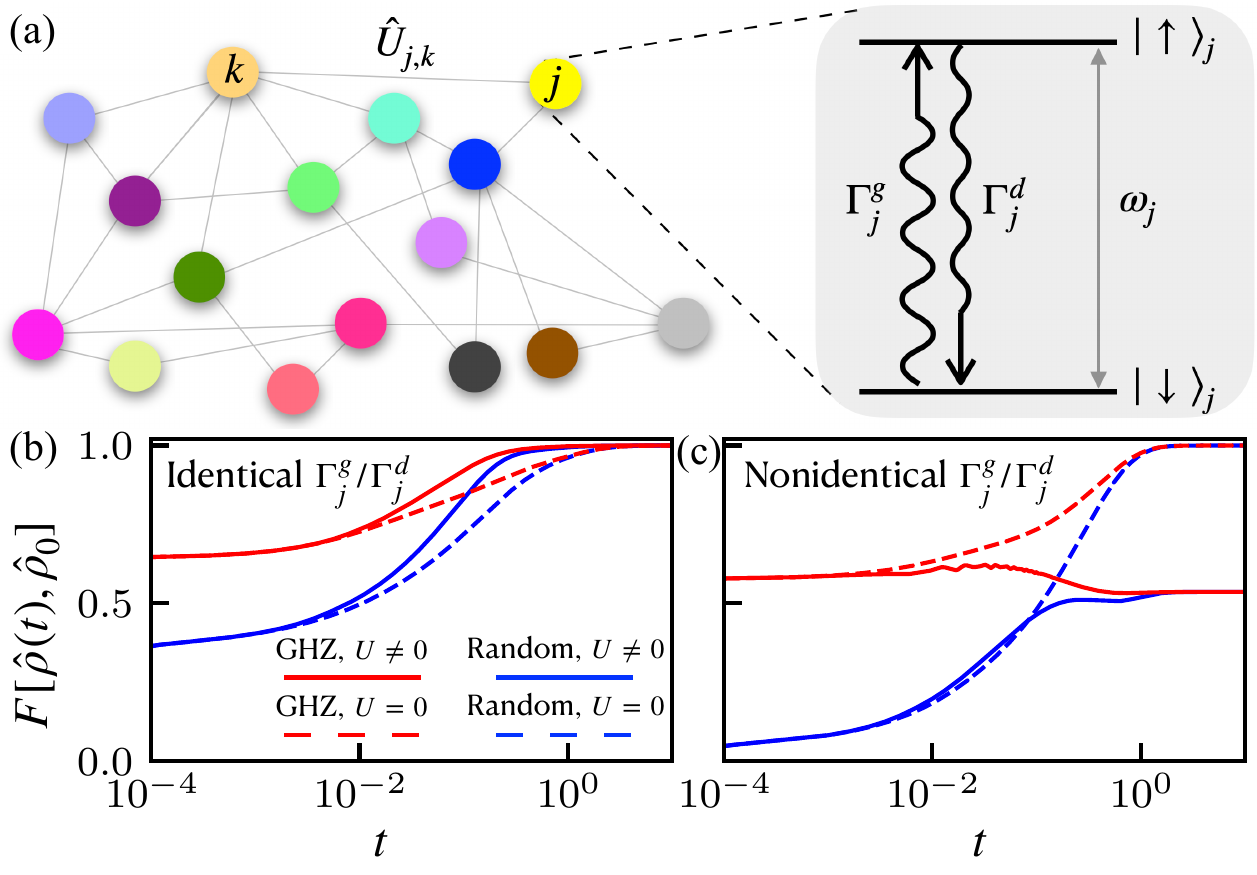}
\caption{\label{fig1} (a) Illustration of the dissipative interacting qubit  system. (b)-(c) Time evolutions of the state fidelity of a six-qubit system with identical $\Gamma^g_j/\Gamma^d_j$ (b) and with nonidentical $\Gamma^g_j/\Gamma^d_j$ (c). The gain rates for the qubits are $\Gamma^{g}_{(1-6)}$=(1,~0.2,~20,~2,~2.5,~0.3) in both (b) and (c). The damping rates $\Gamma^{d}_{(1-6)}$ respectively take (4,~0.8,~80,~8,~10,~1.2) in (b) and (0.5,~4,~80,~1,~0.5,~0.2) in (c). The solid lines  correspond to the all-to-all interacting system with $U^{x}_{jk}=U^{y}_{jk}=80$ and $U^{z}_{jk}=1$. The dashed lines show the corresponding noninteracting cases.}
\end{figure}

{\it System.--- }We consider a collection of  dissipative qubits with both gains and dampings, as illustrated in Fig. \ref{fig1} (a).} The $j$-th qubit is characterized by its frequency $\omega_j$, and  the damping and gain rates, $\Gamma^d_j$ and $\Gamma^g_j$ respectively. The qubits interact with each other via the Heisenberg interaction
\begin{align}\label{eqInter}
\hat{U}=\sum_{j<k}\hat{U}_{jk}=\sum_{j<k}\sum_{\alpha}U_{j,k}^{\alpha}\hat{\sigma}_{j}^{\alpha}\hat{\sigma}_{k}^{\alpha},
\end{align}
where $\hat{\sigma}^{\alpha=x,y,z}_{j}$ are the Pauli operators of the $j$-th qubit. Assuming that the dissipation is Markovian, then the dynamics of this open system is described by the celebrated Lindblad master equation  ($\hbar=1$)
\begin{align}
\label{lindblad}
 \frac{d\hat{\rho}}{dt} = -i\left[\hat{H}_{\text{0}} 
 +  \hat{U}, \hat{\rho}\right] +\frac{1}{2}\sum_j \left (\Gamma^{g}_{j} \mathcal{D}^+[\hat \rho] + \Gamma^{d}_{j} \mathcal{D}^-[\hat \rho] \right)
\end{align}
where $\hat{H}_{\text{0}}=\sum_{j}\omega_{j}\hat{\sigma}_{j}^{z}/2$ , $\hat{\rho}$ is the density matrix of the system, and $\mathcal{D}^\pm[\hat \rho] =  {\hat{\sigma}_j^\pm}\hat{\rho} \hat{\sigma}_j^{\pm\dagger} - \{ \hat{\sigma}_j^{\pm\dagger}  \hat{\sigma}_j^{\pm} , \hat{\rho} \}/2$ with $\hat{\sigma}_{j}^{\pm}\equiv(\hat{\sigma}_{j}^{x}\pm i\hat{\sigma}_{j}^{y})/2$ as the Lindblad operators. The right hand side of Eq.~(\ref{lindblad}) is often written as $\mathcal{L} [\hat \rho]$, where $\mathcal{L}$, referred to the Liouvillian, is the generator of the dynamical semigroup $\{e^{\mathcal{L}t},t\ge0\}$ governing the evolution of the density matrix. The semigroup breaks the time reversibility and evolves the system inevitably to certain steady states $\hat{\rho}_{ss} \equiv \hat\rho(t=+\infty)$, which correspond to the eigenstates of the Liouvillian $\mathcal{L}$ with zero eigenvalues, i.e., $\mathcal{L}[\hat \rho_{ss}] = 0$. Such steady states are the main subjects of study in this paper.

{{\it{No-go theorem.---}} We now state the no-go theorem. {\it Under the conditions that $\Gamma_j^g/\Gamma_j^d$ are the same for all the qubits and the interaction is of the XXZ form {\color{black}($U_{jk}^x = U_{jk}^y$)}, the system governed by Eq.~(\ref{lindblad}) always evolves to an unique steady state in which the correlations between the qubits are completely absent, i.e.,  
\begin{align}
\langle \hat\sigma^\alpha_j \hat\sigma^\beta_k \rangle - \langle \hat\sigma^\alpha_j \rangle\langle \hat\sigma^\beta_k \rangle= 0
\label{nogo}
\end{align}
for any $j\neq k$, where the average is performed with respect to the density matrix of the steady state.} 

{\color{black}Before proving the no-go theorem, we point out three remarkable properties it implies about the dissipated qubit system described by Eq.~(\ref{lindblad}). The first is that strong interactions between the qubits do not necessarily imply strong correlations between them in the steady state.  Secondly, the absence of correlation in the steady state holds for any initial states of the system. Thirdly, there is no constraint on the topology structure of the system to hold the no-go theorem. In other words, this theorem provides a scenario for a system with an arbitrary topology structure, where regardless how strong the interactions are and how strongly correlated the system initially is, no correlations can be established in the end of dynamic evolution.}

{\color{black}We prove the no-go theorem in two steps. First we show that only one steady state exists for the dissipated qubit system given any specific set of parameters.
The Hilbert space of our system is $\mathcal{H}=\otimes_{j} \mathcal{H}_{j}$, where $\mathcal{H}_{j}$ is the local physical Hilbert space of $j$th qubit associated with the generators $\{\hat{I}_{j},\hat{\sigma}^{x}_{j},\hat{\sigma}^{y}_{j},\hat{\sigma}^{z}_{j}\}$. As shown in the Supplemental Material (S.M.) I~\cite{suppmaterials}}, the operators in $\mathcal{H}$ can be generated, through multiplication and addition, by the collection of the jump operators of all qubits, which are the local spin flip down and up operators $\{\sigma_{j}^{\pm}\}$. According to the theorem discussed in Refs.~\cite{evans_irreducible_1977,Frigerio_1978,Prosen_2012,Prosen_2015}, this fact results in the uniqueness of steady state of Eq. (\ref{lindblad})  . 

{\color{black}With the uniqueness of the steady state established}, we now show that the density matrix corresponding to the steady state of the same system in the absence of interactions is a solution to the steady state equation $\mathcal{L}[\hat \rho] = 0$ under the aforementioned conditions. 
{\color{black} For a single qubit, this equation is easily solved and one finds an unique steady state $\hat \rho_{j,0} = (\hat{I}_{j}+{m}_{j}^{z}{{\hat{\sigma}}}^{z}_{j})/2$ for the $j$-th qubit, characterized by the magnetization ${m}_{j}^{z}=1-2\left(\Gamma_{j}^{g}/\Gamma_{j}^{d}+1\right)^{-1}$\cite{PRA2020two,zhang2022}. In the absence of interactions, the steady state density matrix of the many-qubit system is simply
\begin{eqnarray}\label{rho0}
\hat{\rho}_{0}=\prod_{j}\hat{\rho}_{j,0}=\prod_{j}(\hat{I}_{j}+{m}_{j}^{z}{{\hat{\sigma}}}^{z}_{j})/2,
\end{eqnarray}
where $\prod$ denotes the matrix direct products of the sequence of $\hat \rho_{j,0}$. Substituting this expression into the Liouvillian $\mathcal{L}$ of Eq.~(\ref{lindblad}) we obtain
\begin{align}
\mathcal{L}[\hat\rho_0] = [\hat U, \hat \rho_0] = -\frac{1}{2}\sum_{j<k}\left(\hat{\mathcal{M}}_{jk}+\hat{\mathcal{U}}_{jk} \right)\prod_{l \neq j,k}\hat{\rho}_{l,0},
\end{align}
where $ \hat{\mathcal{M}}_{jk}=(m^z_{j} -m^z_{k})(U_{jk}^{x}+U_{jk}^{y})(\hat{\sigma}_{j}^{+}\hat{\sigma}_{k}^{-}-\hat{\sigma}_{j}^{-}\hat{\sigma}_{k}^{+})$ and $ \hat{\mathcal{U}}_{jk}=(m^z_{j} +m^z_{k})(U_{jk}^{x}-U_{jk}^{y})(\hat{\sigma}_{j}^{+}\hat{\sigma}_{k}^{+}-\hat{\sigma}_{j}^{-}\hat{\sigma}_{k}^{-})$. From this expression we see that $\mathcal{L}[\hat \rho_0] = 0$ if  $m_j^z = m_k^z$, or equivalently $\Gamma_j^g/\Gamma_j^d = \Gamma_k^g/\Gamma_k^d$, and $U_{jk}^{x}=U_{jk}^{y} $ for all $j\neq k$. Combined with the uniqueness of the steady state proved earlier, this means that the density matrix  $\hat\rho_0$ given in Eq.~(\ref{rho0}) is the exact and unique steady state of the interacting qubit system under the conditions of identical gain-to-damping ratios and the Heisenberg XXZ interaction. The no-go theorem, i.e., the absence of correlation between any two qubits in the steady state, then follows immediately from Eq.~(\ref{rho0}). We note that the magnetization $m_j^z = 0$ when $\Gamma_j^d = \Gamma_j^g$ and, as a result,  the no-go theorem holds even for a general anisotropic Heisenberg XYZ interaction in this special case. 

{\color{black}We numerically demonstrate the no-go theorem for a six-qubit system through the time evolution of the fidelity between $\hat \rho(t)$ and $\hat \rho_{0}$, defined as $F[\hat \rho(t),\hat \rho_{0}]=\Tr[\sqrt{\hat\rho(t)^{1/2}\hat\rho_{0}\hat\rho(t)^{1/2}}]$ \cite{nielsen2000quantum}, as shown in Fig.~\ref{fig1}(b-c).} To test the uniqueness of the steady state we choose two  initial states for the evolution, the maximally entangled Greenberger–Horne–Zeilinger (GHZ) state (red lines) and a random state (blue lines). For {\color{black}identical $\Gamma_j^g/\Gamma_j^g$} they all evolve to the unique steady state $\hat\rho_0$, as clearly indicated by $F[\hat\rho(+\infty),\rho_{0}]=1$ (see Fig.~\ref{fig1}(b)); this is to be contrasted with the case of {\color{black}nonidentical $\Gamma_j^g/\Gamma_j^g$}, where they evolve to some other unique steady state $\hat \rho(+\infty) \neq \hat \rho_0$ (see Fig.~\ref{fig1}(c)).  We also compare the evolutions of the system with (solid lines) and without (dashed lines) interactions. It is important to note that for identical $m_j^z$ even though the presence of interactions does not alter the final steady state,  it changes the evolution trajectory. We have also numerically confirmed the no-go theorem by directly computing the evolution of various correlations ({\color{black} see S.M. II \cite{suppmaterials}).}  
 
For the rest of the paper we turn to the application of the no-go theorem in the context of quantum synchronization of qubits, where both dissipation and spin correlation are indispensable ingredients.

{\em Quantum synchronization blockade.--- }{\color{black}  As the first application, we show that our no-go theorem can shed light on a novel blockade effect in quantum synchronization of qubits. Here synchronization refers to the tendency of weakly coupled self-sustained oscillators to adjust their rhythm and oscillate at the same frequency with locked phases 
. A central concept of this phenomenon is the limit cycle, a phase-symmetric state to which the oscillator always returns after a weak perturbation. In the case of a dissipated qubit $j$, the limit cycle is represented by its steady state $\hat \rho_{j,0}$.  In classical synchronization, two oscillators can be synchronized only if they possess similar or identical limit cycles. In quantum regime, however, synchronization is suppressed between two identical oscillators, as is found in dissipated Kerr-anharmonic oscillators as well as {\color{black}in larger spins~\cite{PRL2017kerr,PRA2018QSB,Solanki2022}}. We shall see that such a blockade effect also manifests in the qubit system. To quantify the degree of synchronization between the $j$th and $k$th qubits, we follow Ref.~\cite{PRL2018QN} and define the $S$-function~\cite{PRL2018,PRA2019tribit,PRL2020exp,PRA2020two,zhang2022}, $S_{jk}(\phi)=-(2\pi)^{-1} + \int_0^{2\pi}{d}\phi_{k}\int_0^{\pi}{d}\theta_{j}\int_0^{\pi} d\theta_k\sin\theta_{j}\sin\theta_k Q_{jk}({\theta_j},\theta_k,\phi+\phi_k,{\phi_k})$, where $\phi \equiv \phi_j - \phi_k$. Here $Q_{jk} \equiv \bra{\bm{\phi},\bm{\theta}}\hat{\rho}_{jk}\ket{\bm{\phi},\bm{\theta}}/(2\pi)^{2}$ is the Husimi $Q$ function, where $\hat \rho_{jk} $ is the reduced density matrix of the $j$-th and $k$-th qubits, and $\ket{\bm{\phi},\bm{\theta}}\equiv\ket{\phi_{j},\theta_{j}}\bigotimes\ket{\phi_{k},\theta_{k}}$ is the product of coherent spin states, $\ket{\phi_{j},\theta_{j}}=\exp(-i\phi_{j}\hat{\sigma}^{z}_{j}/2)\exp(-i\theta_{j}\hat{\sigma}^{y}_{j}/2)\ket{\uparrow}_{j}$.  It is insightful to rewrite $S_{jk}(\phi)$ as {\color{black}(see the derivation in S.M. III \cite{suppmaterials})}
\begin{eqnarray}\label{S12qubit}
 S_{jk}(\phi)=\frac{\pi}{16}\vert\langle\hat{\sigma}_{j}^{+}\hat{\sigma}_{k}^{-}\rangle\vert \cos(\phi  - \phi^{(0)}),
\end{eqnarray}
where $\phi^{(0)}$ is the phase of the qubit flip-flop correlation function  $\langle\hat{\sigma}_{j}^{+}\hat{\sigma}_{k}^{-}\rangle$.   This expression  shows the tendency of the two qubits to phase lock to $\phi^{(0)}$ since at this phase value the synchronization measure reaches the maximum value $S_{jk}^{\rm max} = \frac {\pi}{16} \vert\langle\hat{\sigma}_{j}^{+}\hat{\sigma}_{k}^{-}\rangle\vert$.

\begin{figure}[t]
\centering
\includegraphics[width=8.6cm ]{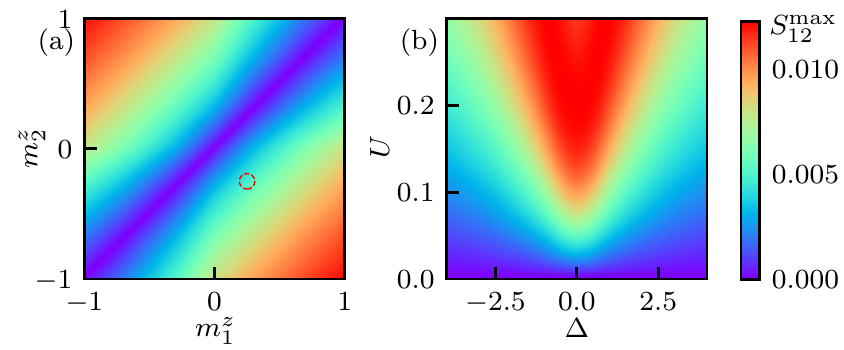}
\caption{\label{fig2} Quantum synchronization of two qubits. For $m^{z}_{j}\geq0~(<0)$, we fix $\Gamma_{j}^{g}=1~(\Gamma_{j}^{d}=1)$. (a) Quantum synchronization blockade with isotropic interaction $U_{12}^{\alpha}=U=1~(\alpha=x,y,z)$ and $\Delta=0$. (b) Arnold tongue corresponding to the circled point in (a), where $m_{1}^{z}=-m_{2}^{z}=1/4$. }
\end{figure}

For a system of two qubits {\color{black}with XXZ interaction ($U_{12}^{x}=U_{12}^{y}=U$),} the spin flip-flop correlation function can be analytically obtained as
\begin{eqnarray}\label{Smax1}
\langle\hat{\sigma}_{1}^{+}\hat{\sigma}_{2}^{-}\rangle= \frac{4U
\Gamma_{1}\Gamma_{2}(m_{1}^{z}-m_{2}^{z}) (4\Delta-i\Gamma)}{
64\Gamma^{2}U^{2}+\Gamma_{1}\Gamma_{2}(\Gamma^{2}+16\Delta^{2})},~~
\end{eqnarray}
where $\Delta = \omega_1 - \omega_2$, $\Gamma_{j}=\Gamma_{j}^{d}+\Gamma_{j}^{g}$ and $\Gamma=\sum_{j=1}^{N}\Gamma_{j}$.  From this expression we find that the synchronization between the two qubits is completely suppressed when $m_1^z = m_2^z$, i.e, the synchronization is blockaded when the qubits have identical limit cycles.  Furthermore, the maximum synchronization occurs when the magnetizations of the two qubits are opposite to each other. These features can be clearly seen in Fig.~\ref{fig2}(a) where we plot $S_{jk}^{\rm max}$ as a function of $m_1^z$ and $m_2^z$ for $\Delta= 0$.  In Fig.~\ref{fig2}(b) we also show one of the hallmarks of synchronization, the formation of the so-called Arnold tongue for the synchronization measure when the frequencies of the qubits are different.  Now, we have demonstrated the synchronization blockade by explicit calculations of the spin correlation functions.  In view of Eqs.~(\ref{nogo}) and (\ref{S12qubit}), we see that this blockade effect follows naturally from our no-go theorem.  In fact, the no-go theorem indicates that the blockade effect persists for multiple qubit systems as long as all the qubits have identical limit cycles. We have numerically verified this by a direct calculation of Eq.~(\ref{S12qubit}) for systems up to six qubits. 

{\em Synchronization of complex qubit networks---} {\color{black} Quantum synchronization blockade is an exact, but in a way adverse, consequence of the no-go theorem. This is because in a qubit network we often wish to strengthen, rather than to weaken the synchronization of the qubits. Take for example the case of quantum memory \cite{zhong2015optically,PRX2020}. The qubit for storing quantum information interacts with many other qubits in the environment and this spin-spin interaction is the most important cause for its decoherence. By synchronizing the environmental qubits to the memory qubit through dissipation engineering, the coherent time of the latter may be increased significantly~\cite{lohe2010quantum,zhang2022}.  However, the general problem of synchronizing multiple qubits quickly becomes intractable as the number of qubits increases. The main difficulty here is that the environmental qubits not only interact with the memory qubit but also interact among themselves, namely the qubits form an all-to-all network. An illustration of such a network is shown in the inset of Fig.~\ref{fig3}(a), where the memory qubit is labeled by $j=1$.  

Here we show that the no-go theorem can be utilized to reduce the complexity of the network, thereby providing a scheme to optimize the synchronization.  The key observation from the no-go theorem is that when the qubits have the identical gain-to-damping ratio, or equivalently the identical magnetization, the spin-spin interactions are as if absent in determining the steady state. This suggests that if we tune the magnetization of the environmental qubits to be the same, the effect of the spin-spin interactions among them can be effectively {\color{black}decoupled}. As a result, the all-to-all qubit network under this condition can be reduced  to a much simpler one-to-all network shown in the inset of Fig.~\ref{fig3}(b).  This argument is of course not completely rigorous because the no-go theorem applies only to the situation where {\it all qubits} have the same magnetization; here we allow the memory qubit to have a different magnetization from that of the environmental qubits. Nevertheless, as will be demonstrated numerically, the equivalence of these two networks with the specified dissipation engineering is remarkably accurate as far as the synchronization is concerned. To be concrete, we define the measure for the synchronization of the multiple qubit system $S_{t}^{\text{max}}=\sum_{j<k}S_{jk}^{\text{max}}$. {\color{black}We numerically calculate $S_{t}^{\text{max}}$ of a five-qubit system for the all-to-all and the one-to-all networks as respectively shown in Fig.~\ref{fig3}(a-b), where in both networks the environmental qubits have identical magnetizations.} We see that the total synchronization measures for these two networks are almost indistinguishable, providing strong evidence for our previous argument based on the no-go theorem. {\color{black}However}, when the magnetizations of the environmental qubits differ from each other, the two networks are clearly inequivalent, as shown by Fig.~\ref{fig3}(c-d).

\begin{figure}[t]
\centering
\includegraphics[width=8.6cm]{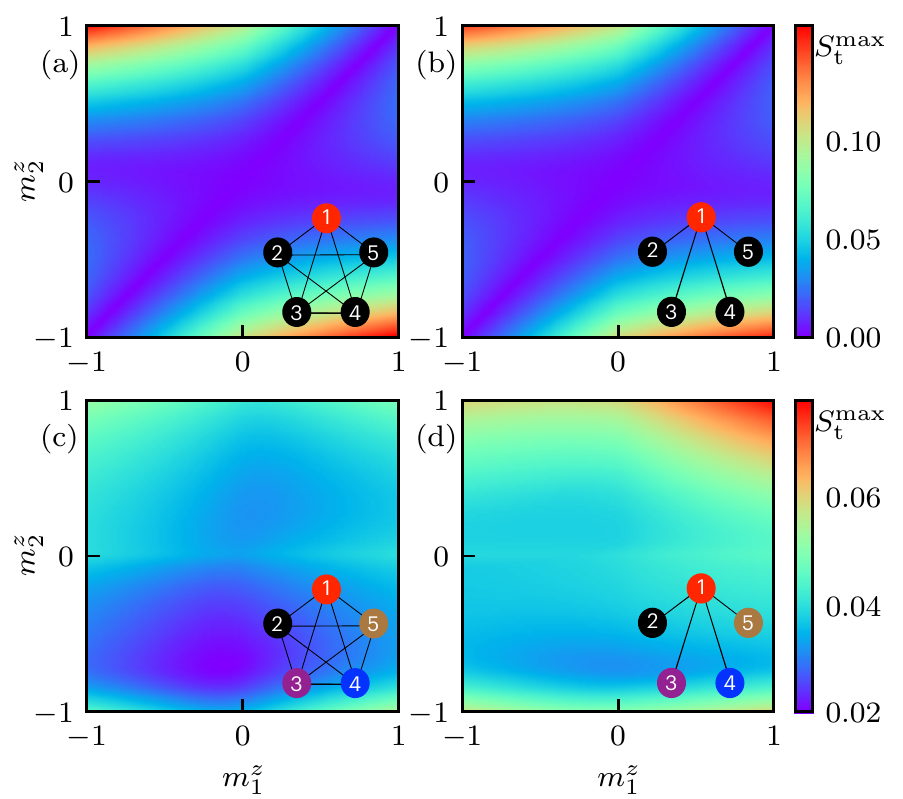}
\caption{\label{fig3} Quantum synchronization of all-to-all (a,c) and one-to-all (b,d) quantum networks. For $m^{z}_{1\leq j\leq 2}\geq$0~($<$0), we fix $\Gamma_{1\leq j\leq 2}^{g(d)}=1$. The interaction parameters are $U_{jk}^{x}=U_{jk}^{y}=2$ in (a-b); and $U_{jk}^{x}=2, U_{jk}^{y}=0.5$ in (c-d).  The $m_j^z$ for $2\le j \le 5$ are identical in (a-b) and are different in (c-d).  {\color{black}More specifically, in (a-b) we have $\Gamma^{g,d}_{(3-5)}/\Gamma^{g,d}_{2}$=(1.5,~2.5,~0.6); in (c-d) we have $\Gamma^{g}_{(3-5)}/\Gamma^{g}_{2}=(0.8,~5,~0.1)$ and $\Gamma^{d}_{(3-5)}/\Gamma^{d}_{2}$=(1.3,~0.9,~1.2). Other parameters are $U_{jk}^{z}=1, \Delta_{jk}=0$.}}
\end{figure}

{\em Experimental implementations. ---} Here we propose two experimental systems to test our no-go theorem. In a recent experimental work by some of us~\cite{zhang2022}, we apply gain, damping and repumping lasers to an eight-level ion and realized a qubit with controlled gain and damping rates. Here, we propose to add an acousto-optic modulator (AOM) to split the lasers to individually manipulate multiple ions \cite{debnath2016demonstration}, as shown in Fig. \ref{fig4}(a). The applied interaction laser globally addresses the ions and couples their qubit states to the collective motional modes. The induced effective interaction is a power-law  generalized Heisenberg XYZ interaction $\hat{V}_{XYZ}=\sum_{j<k}\sum_{\alpha=x,y,z}U_{jk}^{\alpha}\hat{\sigma}_{j}^{\alpha}\hat{\sigma}_{k}^{\alpha}+B\sum_{j}\hat{\sigma}_{j}^{z}${\color{black}, which will reduce to the $XX$ interaction  $\hat{V}_{XX}=\sum_{j<k}U_{jk}(\hat{\sigma}_{j}^{+}\hat{\sigma}_{k}^{-}+\hat{\sigma}_{k}^{-}\hat{\sigma}_{j}^{+})$, under strong magnetic field \cite{jurcevic2014quasiparticle,PRL2009Ion,PRL2004Ion,Science2013,britton2012engineered}.}

\begin{figure}[t]
\centering
\includegraphics[width=8.6cm]{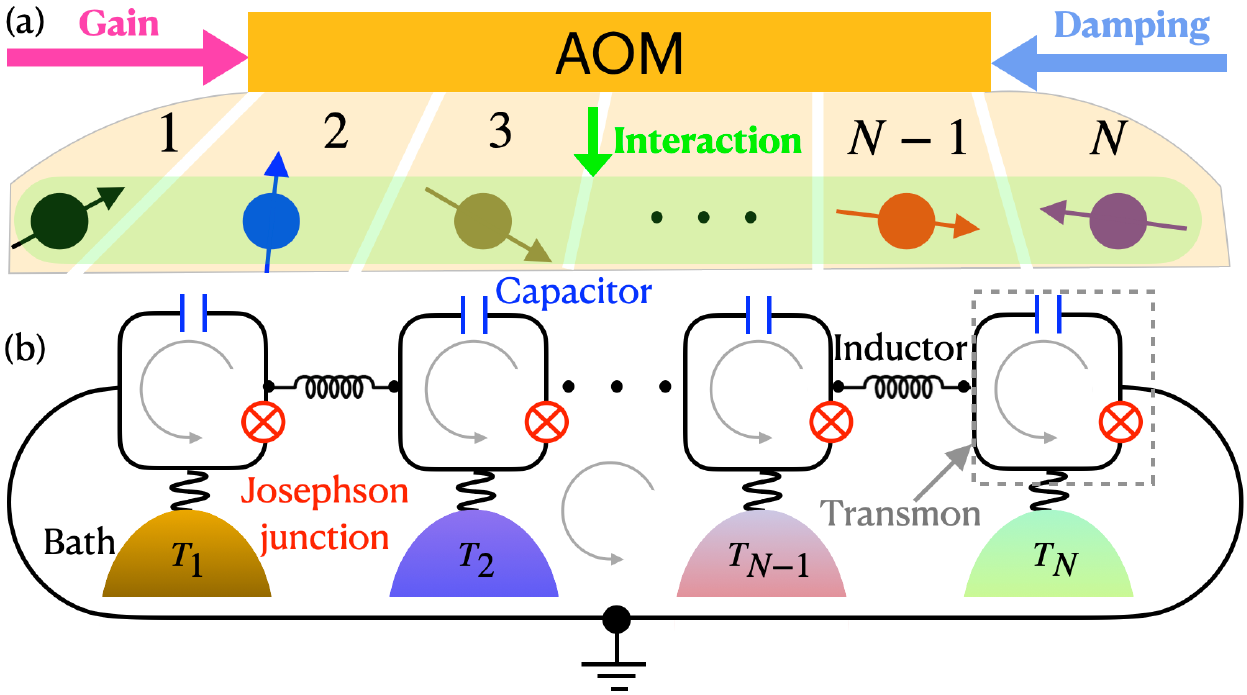}
\caption{\label{fig4} Experimental schemes for the implementation of our model. (a) Trapped ion platform. (b) Superconducting circuit system. }
\end{figure}

A second experimental system is the superconducting circuits as shown in Fig. \ref{fig4}(b). Here we use inductors to couple adjacent transmons-like qubits, each of them comprising one capacitor and Josephson junction. This results in a {\color{black}Heisenberg XX interaction $\hat{V}_{XX}$} between the qubits \cite{PRXQuantum2022,Review2019}. The advantage of this system is that all the elements can be independently controlled. The applied $j$th heat bath independently control the ratio of dissipation rates of $j$th transmons-like qubit, $\Gamma_{j}^{g}/\Gamma_{j}^{d}=1-[1+n_{j}^{\text{B}}(T_{j})]^{-1}$, via tuning the average phonon number $n_{j}^{\text{B}}$, i.e., temperature $T_{j}$ \cite{Bohr_Brask_2015}.

{\em Outlook ---} {\color{black}In addition to the two implications of the no-go theorem discussed in this work, we briefly outline other potential applications and extensions of the theorem. First, the no-go theorem provides a reset mechanism \cite{PRL2013reset,PRA2017reset,PRL2018reset} of qubit systems to arbitrary mixture states in the presence of interaction, independent of qubit positions and topology structure of system. This is useful for experimental studies of the dissipative dynamics, including the dynamical purification phase transition \cite{PRX2020PPT}, for which beginning at a mixture state is often required. Our results can also be generalized to the driven-dissipative qubit system \cite{PRA2015driven,PRL2017driven,Owen2018,PRA2018driven} to further understand the simplification of complex quantum networks chacracterized by the nonequilibrium steady states. Finally, although we have verified that the the no-go theorem does not apply to the spin-1 system under Heisenberg interaction \cite{suppmaterials}, it's still an open question whether similar no-go theorems exist in larger, dissipative spin systems {\color{black} or in other models such as the Hubbard model.} All these problems will be explored in future studies.}
\\

\begin{acknowledgments}
The authors acknowledge Tongxing Yan for the helpful discussions about superconducting qubit circuits. This work is supported by the National Key R\&D Program of China under Grant No. 2022YFA1404103 (Z. W.), the National Natural Science Foundation of China under Grant No. 12104210 (J. J.), No.~11974161 (Z. W.) and No. 12022405  (R. Q.), the China Postdoctoral Science Foundation under Grant No. 2022M711496 (J. J.), No. 2018YFA0306501 (R. Q.), the Beijing Natural Science Foundation under Grant No. Z180013  (R. Q.) and Shenzhen Science and Technology Program under Grant No.~KQTD20200820113010023 (Z. W.).
\end{acknowledgments}
\bibliography{QS_Refs}

\global\long\def\id{\mathbbm{1}}
\global\long\def\ui{\mathbbm{i}}
\global\long\def\ud{\mathrm{d}}

\setcounter{equation}{0} \setcounter{figure}{0}
\setcounter{table}{0} 
\renewcommand{\theparagraph}{\bf}
\renewcommand{\thefigure}{S\arabic{figure}}
\renewcommand{\theequation}{S\arabic{equation}}

\onecolumngrid
\flushbottom
\newpage
\title{Supplementary Material:\\Absence of correlations in dissipative interacting qubits: a no-go theorem}

\maketitle

\section{I. Proof of the uniqueness of the steady state}\label{secI}
We consider the dissipative systems comprising $N$ qubits and its dynamics is characterized by the Lindblad master equation ($\hbar=1$)
\begin{eqnarray}\label{eq1}
 \dot{\hat{\rho}} = \mathcal{L}[\hat{\rho}]=-i[\hat{H}, \hat{\rho}] + \sum_{j} \mathcal{L}_{j}\hat{\rho},
\end{eqnarray}
with the Hamiltonian $\hat{H}$. The dissipations are given by $\mathcal{L}_{j}\hat{\rho}=(\Gamma^{g}_{j} \mathcal{D}[\hat{\sigma}^{+}_{j}] + \Gamma^{d}_{j} \mathcal{D}[\hat{\sigma}_{j}^{-}])\hat{\rho}/2$ with the Lindblad super-operator $\mathcal{D}[\mathcal{\hat{A}}]\hat{\rho} = \mathcal{\hat{A}}\hat{\rho} \mathcal{\hat{A}}^\dagger - \{ \mathcal{\hat{A}}^\dagger \mathcal{\hat{A}}, \hat{\rho} \}/2$ and jump operator $ \mathcal{\hat{A}}$. The time evolution of the open systems in Eq. (\ref{eq1}) is under an semigroup $\{e^{\mathcal{L}t},t\ge0\}$ generated by the Liouvillian $\mathcal{L}$. The semigroup breaks the time reversibility and leads the system to the steady states $\hat{\rho}_{*}$, which is the eigenstates associated with zero eigenvalues of the Liouvillian $\mathcal{L}$, i.e., $\dot{\hat{\rho}}=\mathcal{L}[\hat{\rho}_{*}]=0$. The degeneracy of steady state has important impact of the Liouvillian dynamics. The determine of the uniqueness of steady state, which is the first step for proving our no-go theorem in the main text, has been studied for decades. Here we follow the theorem given by Evans \cite{evans_irreducible_1977} and Frigeiro \cite{Frigerio_1978} to show the uniqueness of the steady state of our system.
The Hilbert space of our system is $\mathcal{H}=\otimes_{j} \mathcal{H}_{j}$, where $\mathcal{H}_{j}$ is the local physical Hilbert space of $j$th qubit associating with the generators $\{\hat{I}_{j},\hat{\sigma}^{x}_{j},\hat{\sigma}^{y}_{j},\hat{\sigma}^{z}_{j}\}$. One theorem given by Evans \cite{evans_irreducible_1977} and Frigeiro \cite{Frigerio_1978} states that the steady state of Eq. (\ref{eq1}) is unique if all the operators in Hilbert space $\mathcal{H}$ can be generated by the Hamiltonian $\hat{H}$,  all the jump operators $\{\mathcal{A}_{j}\}$ and their conjugate operators $\{\mathcal{A}_{j}^{\dagger}\}$ under multiplication and addition. In our system, each qubit here takes the local spin flip down and up operators $\{\sigma_{j}^{\pm};j = 1,2,\cdots, N\}$ as the jump operators, the collection of them obviously can rebuilt the Hilbert space $\mathcal{H}$. For example, the pauli operators and identical operator for $j$th qubit can be generated as following, 
\begin{eqnarray}
\hat{\sigma}_{j}^{x}&=&\hat{\sigma}_{j}^{+}+\hat{\sigma}_{j}^{-},\\
\hat{\sigma}_{j}^{y}&=&(\hat{\sigma}_{j}^{+}-\hat{\sigma}_{j}^{-})/i,\\
\hat{\sigma}_{j}^{z}&=&\hat{\sigma}_{j}^{+}\hat{\sigma}_{j}^{-}-\hat{\sigma}_{j}^{-}\hat{\sigma}_{j}^{+},\\
\hat{I}_{j}&=&\hat{\sigma}_{j}^{+}\hat{\sigma}_{j}^{-}+\hat{\sigma}_{j}^{-}\hat{\sigma}_{j}^{+}.
\end{eqnarray}
Therefore our system possesses a unique steady state. This theorem was also applied to other dissipative Heisenberg spin chain \cite{PRL2011Exact,Prosen_2012,Prosen_2015}.

\section{II. Dynamics of correlations}
Figure 1(b-c) in the main text numerically illustrates the no-go theorem by the fidelity. Here we in further show the results of correlation functions which are defined as the summation of the correlation between qubits,
\begin{eqnarray}
C_{\alpha\beta} &=& \sum_{j< k}\mathrm{Tr}[ \hat{\rho}_{jk}\hat{\sigma}^\alpha_j\hat{\sigma}^\beta_k];\alpha,\beta=x,y,\pm,
\end{eqnarray}
\begin{figure}
\centering
\includegraphics[width=\linewidth]{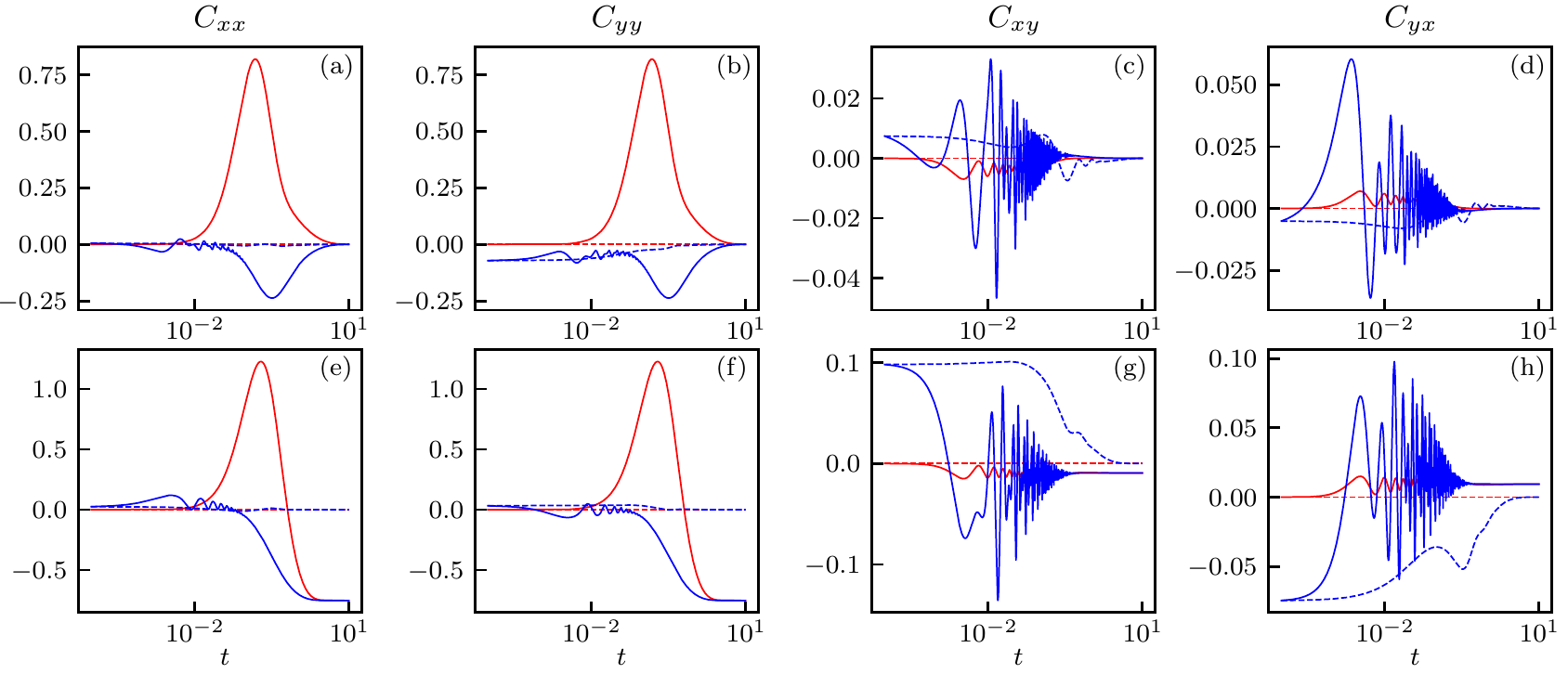}
\caption{(a-d): The correlation functions corresponding
to the cases in Fig. 1(b) in the main text. (e-h): The correlation functions corresponding to the cases in Fig. 1(c) in the main text. The panels of each column share the same $y$-axis.  }
\label{Sfig1}
\end{figure}
where $\hat{\rho}_{jk} = \mathrm{Tr}^{l\neq j, k}[\hat{\rho}]$ is the reduced density matrix operator for the combination of $j$th qubit and $k$th qubit. To understand the correlation functions better, we rewrite the reduced density matrix $\rho_{jk}$ in a general form,
\begin{align}
  \rho_{jk} =
  \begin{pmatrix}
    a & p^{*} & q^{*} & r^{*} \\
    p & b     & s^{*} & u^{*} \\
    q & s     & c     & v^{*} \\
    r & u     & v     & d
  \end{pmatrix},
\end{align}
then we can see the correlation functions,
\begin{eqnarray}
\label{eq:4}
  C_{-+} &=& s^{*} ,\\
  C_{+-} &=& s ,\\
  C_{++} &=& r ,\\
  C_{--}&=& r^{*},\\
  C_{xx} &=& r + s + s^{*} + r^{*}  = 2(\mathrm{Re}[s] + \mathrm{Re}[r]),\\
 C_{yy} &=& - r + s + s^{*} - r^{*}  = 2(\mathrm{Re}[s] - \mathrm{Re}[r]),\\
 C_{xy}&=& i(- r + s - s^{*} + r^{*})  = 2(\mathrm{Im}[r] - \mathrm{Im}[s]),\\
  C_{yx} &=& i(- r - s + s^{*} + r^{*})  = 2(\mathrm{Im}[r] + \mathrm{Im}[s]),
\end{eqnarray}
Therefore, we only consider the correlation functions $\{C_{++}, C_{--}, C_{+-}, C_{-+}\}$, because the correlation functions $\{C_{xx}, C_{xy}, C_{yx}, C_{yy}\}$ contain the same information. Figure \ref{Sfig1} shows the numerical results of those correlation function using the same parameter of Fig. 1(b-c) in the main text.  In Fig. \ref{Sfig1}(a-d), all the curves vanishes at long evolution time and verifies the no-go theorem. As shown in Fig. \ref{Sfig1}(e-h), when the qubits do not have the same limit cycle, the solid lines will away from zero eventually which illustrates the breakdown of the no-go theorem.



\section{III. Derivation of Eq. (6) in the main text}
To derive Eq. (6) in the main text, we first review the quantum sychronization measure for a single dissipated qubit \cite{PRA2020two,zhang2022}. Here we follow \cite{PRL2018,PRL2018QN,PRA2020two} to apply the Husimi-$Q$ function for visualizing the density matrix in phase space,
\begin{align}\label{Q1}
  Q(\theta, \phi) = \frac{1}{2\pi} \langle \theta, \phi| \hat{\rho} |\theta, \phi \rangle
\end{align}
where $|\theta, \phi\rangle = e^{-i\frac{\phi}{2}\hat{\sigma}^z}
e^{-i\frac{\theta}{2}\hat{\sigma}^y} |\uparrow\rangle=e^{-i\frac{\phi}{2}}\cos(\frac{\theta}{2})|\uparrow\rangle+e^{i\frac{\phi}{2}}\sin(\frac{\theta}{2})|\downarrow\rangle$ is spin coherent states. Thus for a single qubit in the spin coherent state $|\theta, \phi\rangle$, the angle $\theta$ relates to its energy, and the angle $\phi$ is the relative phase between two spin states. Thus in quantum synchronization, the angle $\phi$ is the free phase for phase locking. Husimi-$Q$ function $Q(\theta, \phi)$ describe the quasi-probability distribution in the Bloch sphere, where the Bloch vector of qubit $\vec{m} = (m_x , m_y, m_z)$ relates to the density matrix as $\rho = \frac{1}{2}(I + \vec{m}\cdot\vec{\sigma})$. Then the straightforward calculation of Eq. (\ref{Q1}) yields,
\begin{align}\label{Q2}
  Q(\theta, \phi) = \frac{1}{4\pi} (1 + \vec{m}\cdot \vec{n}),
\end{align}
where $\vec{n} = (\sin\theta\cos\phi , \sin\theta\sin\phi , \cos\theta)$ is the direction pointing to the spin coherent state $|\theta, \phi\rangle $ in the Bloch sphere. 

To measure the strength of quantum synchronization, we can integrate $\theta$ of Husimi-$Q$ function $Q(\theta, \phi)$ and define the synchronization measure $S$ function \cite{PRA2020two} as below,
\begin{align}\label{eq:S}
S(\phi) =& -\frac{1}{2\pi} + \int_0^{\pi}  Q(\theta, \phi)\sin\theta\mathrm{d}\theta,\nonumber\\
=& \frac{1}{8}(m_x \cos\phi + m_y\sin\phi),\nonumber\\
=&\frac{1}{4}\left( \mathrm{Re}\langle\sigma^+\rangle\cos\phi
     + \mathrm{Im}\langle\sigma^+\rangle\sin\phi \right).
\end{align}

The synchronization measure $S$ function in Eq. (\ref{eq:S}) for a single qubit can be straightforwardly generalized to two qubits as follows \cite{PRL2018},
\begin{align}\label{srel1}
  S_{\mathrm{rel}}(\phi) =& -\frac{1}{2\pi} + \int_0^{2\pi}\mathrm{d}\phi_2
  \int_0^{\pi}\mathrm{d}\theta_1\int_0^{\pi}\mathrm{d}\theta_2 \cdot \sin\theta_1\sin\theta_2
  Q_{12}(\theta_1, \theta_2, \phi + \phi_2, \phi_2), 
\end{align}
where $\phi = \phi_1 - \phi_2$ is the relative phase.
\begin{align}\label{Q12}
  Q_{12}(\theta_1, \theta_2, \phi + \phi_2, \phi_2) = \frac{1}{(2\pi)^2}\langle \theta_1,\theta_2,\phi_1,\phi_2 |
  \hat{\rho}_{12}| \theta_1,\theta_2,\phi_1,\phi_2\rangle
\end{align}
with$| \theta_1,\theta_2,\phi_1,\phi_2\rangle = | \theta_1,\phi_1\rangle \otimes | \theta_2, \phi_2\rangle$

For any state $ \hat{\rho}_{12}$, we can write it as 
 \begin{align}
  \hat{\rho}_{12}=\sum_{p_j}p_j |\psi^j\rangle \langle\psi^j|,
\end{align}
where $p_j$ is the coefficient for the pure state $|\psi^j\rangle$, the Schmidt decomposition of which is
\begin{align}
|\psi\rangle = \sum_{k}  \lambda_k |u_k\rangle_1 \otimes |v_k\rangle_2,
\end{align}
with the Schmidt coefficients $\lambda_k$. Thus we can decompose any two qubits state $ \hat{\rho}_{12}$ as the sum of tensor products,
\begin{align}  \label{eq:Shmit}
\hat{\rho}_{12} =&  \sum_{p_j, \lambda_k, \lambda_l}
  p_j \lambda_k \lambda_l \left(|u^j_k\rangle_1 \otimes |v^j_k\rangle_2\right)
            \left(\langle u^j_l|_1 \otimes \langle v^j_l|_2\right), \nonumber\\
  =& \sum_{p_j, \lambda_k, \lambda_l}
     p_j \lambda_k  \lambda_l \left(|u^j_k\rangle_1 \langle u^j_l|_1 \right)
     \otimes \left( |v^j_k\rangle_2 \langle v^j_l|_2\right), \nonumber\\
  =& \sum_{p_j, \lambda_k, \lambda_l}p_j \lambda_j  \lambda_l \cdot \hat{\rho}_{1,kl}^j \otimes
     \hat{\rho}_{2,kl}^{j}, \nonumber\\
    =& \sum_w C_w \cdot \hat{\rho}_{1,w} \otimes \hat{\rho}_{2,w},
\end{align}
where $w = \{p_i, \lambda_k,  \lambda_l\}$ , $C_w = p_i \lambda_k  \lambda_l$. Note that $\hat{\rho}_{j,w}$ is a general operator and may not density matrix 
We define the associated general Husimi-$Q$ function as $Q_{jw}(\theta_j,\phi_j)=\langle \theta_1,\phi_1|\hat{\rho}_{1,w}| \theta_1,\phi_1\rangle/2\pi$. Then the related general $S$ function is 
\begin{align}
S_{jw}(\phi) =& -\frac{ \mathrm{Tr}[\rho_{j,w}]}{2\pi} + \int_0^{\pi}   Q_{jw}(\theta, \phi)\sin\theta\mathrm{d}\theta,\nonumber\\
=&\frac{1}{8}\left( \langle\sigma_j^x\rangle_{w}\cos\phi
     + \langle\sigma_j^y\rangle_{w}\sin\phi \right).
\end{align}
Then subsituting Eq. (\ref{eq:Shmit}) into the Husimi-$Q$ function in Eq. (\ref{Q12}), we have 
\begin{align}
Q_{12}(\theta_1, \theta_2, \phi + \phi_2, \phi_2) = \sum_w C_w Q_{1w}(\theta_1, \phi + \phi_2)   Q_{2w}(\theta_2, \phi_2).
\end{align}
By substituting above $Q_{12}$ into $S_{\mathrm{rel}}(\phi)$ in Eq. (\ref{srel1}), we have
\begin{align}\label{eq:S-rel}
S_{\mathrm{rel}}(\phi) =& -\frac{1}{2\pi} + \int_0^{2\pi}\mathrm{d}\phi_2
  \int_0^{\pi}\mathrm{d}\theta_1\int_0^{\pi}\mathrm{d}\theta_2 \cdot \sin\theta_1\sin\theta_2 \sum_{w}C_wQ_{1w}(\theta_1, \phi + \phi_2)   Q_{2w}(\theta_2, \phi_2), \nonumber\\
  =& -\frac{1}{2\pi} + \int_0^{2\pi}\mathrm{d}\phi_2\cdot \sum_{w}C_w
     \left[\frac{\mathrm{Tr}[\rho_{1,w}]}{2\pi} + S_{1w}(\phi + \phi_2)\right]
     \left[\frac{\mathrm{Tr}[\rho_{2,w}]}{2\pi} + S_{2w}(\phi_2)\right], \nonumber\\
  =&\int_0^{2\pi}\mathrm{d}\phi_2\cdot
     \sum_{w}C_w S_{1w}(\phi + \phi_2) S_{2w}(\phi_2), \nonumber\\
   =& \frac{1}{64}\int_0^{2\pi}\mathrm{d}\phi_2\cdot\sum_wC_w\left[ \langle\sigma^{x}_{1}\rangle_w\cos(\phi+\phi_2)
     + \langle\sigma^{y}_{1}\rangle_w\sin(\phi+\phi_{2}) \right]\left[ \langle\sigma^{x}_{2}\rangle_w\cos\phi_{2}
     + \langle\sigma^{y}_{2}\rangle_w\sin\phi_{2} \right],\nonumber\\
      =& \frac{\pi}{64}\sum_wC_w\left[(\langle\sigma^{x}_{1}\rangle_w\langle\sigma^{x}_{2}\rangle_w+\langle\sigma^{y}_{1}\rangle_w\langle\sigma^{y}_{2}\rangle_w)\cos\phi
     + (\langle\sigma^{y}_{1}\rangle_w\langle\sigma^{x}_{2}\rangle_w-\langle\sigma^{x}_{1}\rangle_w\langle\sigma^{y}_{2}\rangle_w)\sin\phi\right],\nonumber\\
           =& \frac{\pi}{64}\sum_wC_w\left[\langle\sigma^{x}_{1}\sigma^{x}_{2}+\sigma^{y}_{1}\sigma^{y}_{2}\rangle_w\cos\phi
     + \langle\sigma^{y}_{1}\sigma^{x}_{2}-\sigma^{x}_{1}\sigma^{y}_{2}\rangle_w)\sin\phi\right],\nonumber\\
            =& \frac{\pi}{64}\left[\langle\sigma^{x}_{1}\sigma^{x}_{2}+\sigma^{y}_{1}\sigma^{y}_{2}\rangle\cos\phi
     + \langle\sigma^{y}_{1}\sigma^{x}_{2}-\sigma^{x}_{1}\sigma^{y}_{2}\rangle)\sin\phi\right],\nonumber\\
  =& \frac{\pi}{16}\left(\mathrm{Re}\langle \sigma_1^+ \sigma_2^-\rangle\cos\phi
     + \mathrm{Im}\langle \sigma_1^+ \sigma_2^-\rangle\sin\phi \right),\nonumber\\
  =&\frac{\pi}{16}\vert\langle\hat{\sigma}_{1}^{+}\hat{\sigma}_{2}^{-}\rangle\vert \cos(\phi  - \phi^{(0)}),
\end{align}
where $\langle \hat{O} \rangle_w = \mathrm{Tr}[\hat{\rho}_{1, w}\otimes \hat{\rho}_{2, w} \hat{O}]$ and we have used the fact that $\int \mathrm{d}\phi S_{jw}(\phi) = 0$ during the integration. Thus we arrive at Eq. (6) of the main text, where $\phi_0 = \arctan(\mathrm{Im}\langle \hat{\sigma}_1^+ \hat{\sigma}_2^- \rangle/\mathrm{Re}\langle \hat{\sigma}_1^+ \hat{\sigma}_2^- \rangle)$ is the synchronized phase defined with spin flip-flop correlation $\langle \hat{\sigma}_1^+ \hat{\sigma}_2^- \rangle$. 

\section{IV. The absence of no-go theorem in spin-1 system}
In this section, we explain that the spin-1 system studied in \cite{PRL2018,PRL2018QN} does not possess the no-go theorem discussed in the main text for qubits. For a single spin-1 system, we label the three energy levels as $|1, 1\rangle$ , $|1, 0\rangle$ and $|1, -1\rangle$ .
The spin in \cite{PRL2018,PRL2018QN} is incoherently jumping from two side states $|1, \pm1\rangle$ to the target state $|1, 0\rangle$. This dissipation scheme drives each spin into the following
steady state
\begin{align}
\rho_{\mathrm{LC}} =
  \begin{pmatrix}
    0 & 0 & 0 \\
    0 & 1 & 0 \\
    0 & 0 & 0 \\
  \end{pmatrix},
\end{align}
which is just pure state $|1, 0\rangle$.  This state is the limit cycle state for spin-1 system \cite{PRL2018,PRL2018QN}.

Now we consider Heisenberg XYZ interaction as below
\begin{align}
  \hat{U}= U^x \hat{J}^x_1\hat{J}^x_2
  + U^y \hat{J}^y_1\hat{J}^y_2
  + U^z \hat{J}^z_1\hat{J}^z_2,
\end{align}
where $\hat{J}^{\alpha=x,y,z}_{j=1,2}$ is the angular momentum operator which
satisfy the commutation relation
$[\hat{J}_j^{\alpha}, \hat{J}_j^{\beta}] = \mathrm{i}\epsilon_{\alpha,\beta,\gamma}\hat{J}_j^{\gamma}$. The Lindblad master equation ($\hbar=1$) for two spin-1 is
\begin{eqnarray} \label{spin1lmeq}
 \frac{d\hat{\rho}_{12}}{dt} =-i\left[\sum_{j=1,2}\frac{\omega_{j}}{2}\hat{J}_{j}^{z}+\sum_{\alpha=x,y,z}U^{\alpha}\hat{J}_{1}^{\alpha}\hat{J}_{2}^{\alpha}, \hat{\rho}_{12}\right] +\frac{1}{2} \sum_{j=1,2} \left(\gamma^{g}_{j} \mathcal{D}[\hat{J}^{+}_{j}] + \gamma^{d}_{j} \mathcal{D}[\hat{J}_{j}^{-}]\right)\hat{\rho}_{12}.
\end{eqnarray}
Similar to the qubits system studied in the main text, the steady state of Eq. (\ref{spin1lmeq}) is the product state of limit cycle state $\rho_{\mathrm{LC}} $ for each spin-1 system when turn off the interaction. Using the theorem given by Evans \cite{evans_irreducible_1977} and Frigeiro \cite{Frigerio_1978} as mentioned in Sec. \ref{secI}, it's straightforwardly to prove that the steady state in spin-1 system is unique.
Then we follow the analysis as shown in the Eq. (5) in main text to calculate the commutator below,
\begin{align}\label{eq12}
 [U, \rho_{1, \mathrm{LC}}\otimes\rho_{2, \mathrm{LC}}] =
\begin{pmatrix}
 0 & 0 & 0 & 0 & U^y-U^x & 0 & 0 & 0 & 0 \\
 0 & 0 & 0 & 0 & 0 & 0 & 0 & 0 & 0 \\
 0 & 0 & 0 & 0 & -U^x-U^y & 0 & 0 & 0 & 0 \\
 0 & 0 & 0 & 0 & 0 & 0 & 0 & 0 & 0 \\
 U^x-U^y & 0 & U^x+U^y & 0 & 0 & 0 & U^x+U^y & 0 & U^x-U^y \\
 0 & 0 & 0 & 0 & 0 & 0 & 0 & 0 & 0 \\
 0 & 0 & 0 & 0 & -U^x-U^y & 0 & 0 & 0 & 0 \\
 0 & 0 & 0 & 0 & 0 & 0 & 0 & 0 & 0 \\
 0 & 0 & 0 & 0 & U^y-U^x & 0 & 0 & 0 & 0 \\
\end{pmatrix}.
\end{align}
It's very clear that Eq. (\ref{eq12}) can not vanish unless $U_y = U_x = 0$. That is the no-go theorem studied in the main text is forbidden for the spin-1 system in \cite{PRL2018,PRL2018QN}

\end{document}